\documentstyle[epsfig]{aipproc}

\begin{document}
\title{The Dynamics of Non-Crystalline Silica:\\ Insight from Molecular
Dynamics Computer Simulations}

\author{Walter Kob, J\"urgen Horbach, and Kurt Binder}
\address{Institute of Physics, Johannes Gutenberg University,
Staudinger Weg 7, D-55099 Mainz, Germany}

\maketitle

\begin{abstract}
Using a large scale molecular dynamics computer simulation we
investigate the dynamics of a supercooled melt of SiO$_2$. We find that
with increasing temperature the temperature dependence of the diffusion
constants crosses over from an Arrhenius-law, with activation energies
close to the experimental values, to a power-law dependence. We show
that this crossover is related to the fact that at low temperatures
the dynamics of the ions is dominated by hopping processes, whereas
at high temperatures it shows the continuous flow-like motion
proposed by the ideal version of mode-coupling theory (MCT). Finally we
show that at low temperatures the dynamics of the system in the
$\beta-$relaxation regime obeys the factorization property, in
agreement with MCT.

\end{abstract}

\section*{Introduction}
In the last few years a significant advance in our understanding of the
dynamics of glass-forming liquids has occurred. By now it is well
established that features like, e.g., the cage-effect, the slow dynamics in
the $\beta-$relaxation regime, and the stretching of the relaxation
curves in the $\alpha-$relaxation regime, are characteristic for the
dynamics of supercooled liquids, and that many of these features are
described well, or have even been predicted, by mode-coupling theory
(MCT)~\cite{mct}. For {\it fragile} glass-formers~\cite{angell85} there
are even systems for which MCT describes the relaxation dynamics
correctly not only on a qualitative level, but even on a quantitative
one, such as colloidal particles or Lennard-Jones
systems~\cite{colloids,nauroth97,gleim98}. The dynamics of {\it strong}
glass-formers is, however, understood in much less detail since it is,
e.g., not yet clear whether the features predicted by MCT to be present
in the $\beta-$relaxation regime, such as the critical decay or the von
Schweidler law, do not exist in these system or are just obscured by
other dynamical features, such as the boson-peak. In the present paper
we therefore present some results of a large scale computer simulation
in which we investigated the dynamics of a silica melt, the prototype
of a strong glass former.

\section*{Model and Details of the Simulation}
The silica model we study has been proposed by van Beest {\it et al.}
(BKS)~\cite{beest90} and has been found in the past to reproduce well
various structural properties of real silica~\cite{vollmayr96}. In this
potential the interaction between the ions is given by
\begin{equation}
\phi(r_{ij})=\frac{q_i q_j e^2}{r_{ij}}+A_{ij}\exp(-B_{ij}r_{ij})-
\frac{C_{ij}}{r_{ij}^6}\quad .
\label{eq1}
\end{equation}
Here $r_{ij}$ is the distance between ions $i$ and $j$, and the values
of the partial charges $q_i$ and the constants $A_{ij}$, $B_{ij}$, and
$C_{ij}$ can be found in Refs.~\cite{beest90,vollmayr96}. The
simulations have been done at constant volume using 8016 ions and a box
size of 48.37\AA, thus at a density of 2.37g/cm$^3$, close to the
experimental value of the density. (Such a large system was needed to
avoid finite size effect in the dynamics~\cite{horbach96}.) Using Ewald
sums to evaluate the Coulombic part of the potential, the equations of
motion have been integrated with the velocity form of the Verlet
algorithm with a time step of 1.6~fs. The temperatures investigated were
6100~K, 5200~K, 4700~K, 4300~K, 4000~K, 3760~K, 3580~K, 3400~K, 3250~K,
3100~K, 3000~K, 2900~K and 2750~K, and in order to improve the
statistics of the results we averaged at each temperature over two
independent runs. At each temperature the system was first equilibrated
over a time span which significantly exceeded the typical relaxation
time of the system at this temperature. At the lowest temperature the
runs had 13 million time steps, which is equivalent to about 20~ns of
real time. The total CPU time used for this temperature was 13 years
of single processor time on a CRAY-T3E.

\section*{Results}
The simplest quantity which characterizes the dynamics of a system is
$\langle r^2(t)\rangle$, the mean squared displacement (MSD) of a
tagged particle, which is calculated from the positions $\vec{r}_j(t)$
by means of

\begin{equation}
\langle r^2(t) \rangle =\frac{1}{N_{\alpha}} \sum_{j=1}^{N_{\alpha}}
\langle |{\vec{r}}_j(t)-{\vec{r}}_j(0)|^2\rangle\quad \mbox{with } \alpha
\in \{{\rm Si,O}\}.
\label{eq2}
\end{equation}

\begin{figure}[t!] 
\centerline{\epsfig{file=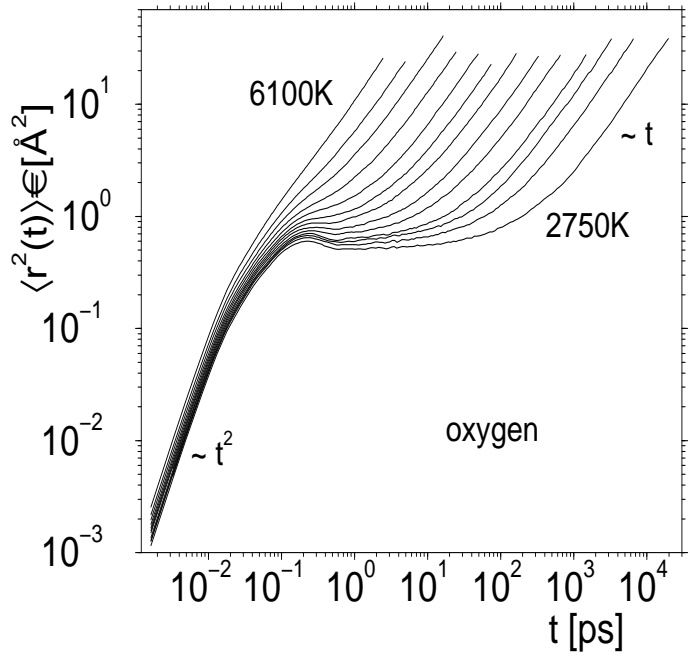,height=3.8in,width=3.8in}}
\vspace{5pt}
\caption{Time dependence of the mean squared displacement for the
oxygen atoms for all temperatures investigated.}
\label{fig1}
\end{figure}

In Fig.~\ref{fig1} we show the time dependence of $\langle
r^2(t)\rangle$ for the oxygen atoms for all temperatures investigated.
(Qualitatively similar curves are found for the silicon
atoms~\cite{horbach98a}.) From this figure we see that at high
temperatures the MSD crosses
over from the ballistic motion of the particles, $\langle
r^2(t)\rangle \propto t^2$, to the diffusive motion, $\langle
r^2(t)\rangle\propto t$. At low temperatures we see in addition to
these two regimes also a third one at intermediate time scales, in which
the MSD stays essentially constant over 2-3 decades in time. In this
time regime the particle is trapped in the cage formed by the atoms
that surround it, and it takes the particle a long time before it is
able to leave this cage. One of the main features of MCT is to give a
self-consistent description of this motion, and below we will discuss
some of these predictions. (Self-consistency is needed since the
particles forming the cage are trapped by {\it their} neighbors.)

\begin{figure}[t!]
\centerline{\epsfig{file=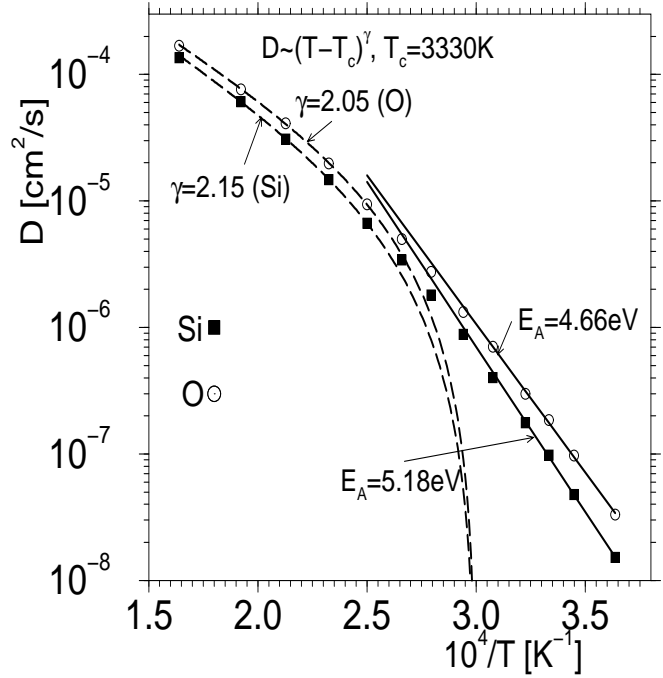,height=3.8in,width=3.8in}}
\vspace{5pt}
\caption{Arrhenius plot of the diffusion constants. The solid lines
are Arrhenius fits to our data at low temperatures. The dashed lines
are power-law fits to our data at high temperatures with an exponent
$\gamma$.}
\label{fig2}
\end{figure}

Using the Einstein relation $D=\lim_{t\to \infty} \langle
r^2(t)\rangle/6t$, it is simple to calculate the diffusion
constant $D$ from the MSD. The temperature dependence of $D$ is shown in
Fig.~\ref{fig2} for $D_{\rm Si}$ and $D_{\rm O}$ in an Arrhenius plot.
We see that at low temperatures the diffusion constants show the
Arrhenius dependence expected for a strong glass-former. The activation
energies found, 5.18~eV and 4.66~eV for silicon and oxygen,
respectively, are in very good agreement with the experimental values
of Br\'ebec {\it et al.} (silicon: 6~eV) and Mikkelsen (oxygen:
4.7~eV)~\cite{brebec_mikkelsen}. Since a similar good agreement has
been found for the temperature dependence of the
viscosity~\cite{horbach98d} we thus conclude that the present model of
silica is not only able to reproduce reliably the static properties of
amorphous silica~\cite{vollmayr96}, but also the dynamic ones.

From the figure we also see that at high temperatures significant
deviations from the Arrhenius law are observed. Such deviations are
often found in fragile glass-formers and have been explained by
means of MCT, which predicts that in supercooled systems there exists a
critical temperature $T_c$ in the vicinity of which transport
coefficients, such as the diffusion constant or the $\alpha-$relaxation
time $\tau$, show a power-law dependence on temperature, i.e.
\begin{equation}
D \propto \tau^{-1} \propto (T-T_c)^{\gamma} \quad,
\label{eq3}
\end{equation}
where the exponent $\gamma$ is a system universal parameter, i.e. is
independent of the transport coefficient. Motivated by this theoretical
prediction we thus have attempted to fit the temperature dependence of
the diffusion constants with such a power-law and have included the
results of these fits in Fig.~\ref{fig2} as well (dashed lines). Note
that in these fits the critical temperature $T_c$ was {\it not} used as
a fit parameter but fixed to 3330~K, a value we obtained from fits to
the relaxation times $\tau$ of the intermediate scattering function
for different values of the wave-vector $q$
(which will be discussed below). From the figure we can see that the so
obtained fits reproduce the data very well over a temperature range in
which the diffusion constants change by about 1.5 decades. This range
is significantly smaller than the one typically found in fragile
glass-formers (3-4 decades~\cite{kob_lj}) and it is often argued that
such a small range is related to the dominant presence of activated
processes, the so-called hopping processes, which are taken into
account only in the ``extended'' version of MCT~\cite{mct_hopping} and
not in the ``ideal'' version~\cite{mct}.

One way to see the mentioned hopping processes directly is to study
the self part of the van Hove correlation function, $G_s(r,t)$, which
is given by

\begin{equation}
G_s^{\alpha}(r,t)=\frac{1}{N_{\alpha}}\sum_{i=1}^{N_\alpha}
\langle \delta(r-|\vec{r}_i(t)-\vec{r}_i(0)|) \rangle \qquad
\alpha \in \{{\rm Si,O}\}\quad.
\label{eq4}
\end{equation}

\begin{figure}[t!]
\centerline{\epsfig{file=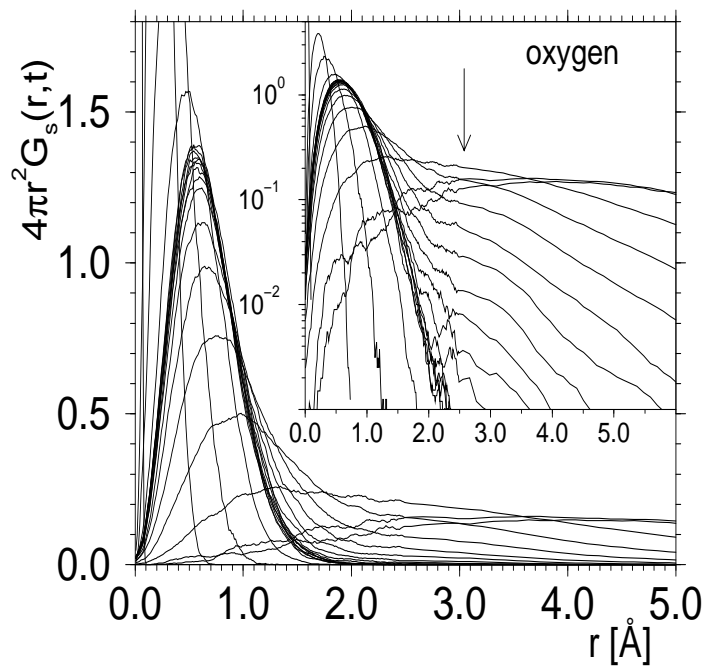,height=3.8in,width=3.8in}}
\vspace{5pt}
\caption{$r-$dependence of the self part of the van Hove correlation
function for the oxygen atoms. The times are 0.16~ps, 1.44~ps and then
increase by a factor of about 2.3. The last curves are 15.2~ns and 19.2~ns .
Inset, the same curves in a lin-log plot. The arrow marks the location
of the first peak in the radial distribution function between two
oxygen atoms.}

\label{fig3}
\end{figure}

Thus $4\pi r^2G_s^{\alpha}(r,t)$ is the probability to find a particle
at time $t$ at a distance $r$ away from the place it was at $t=0$. In
Fig.~\ref{fig3} we show the $r-$dependence of $4\pi r^2G_s^{O}(r,t)$
for different times for the lowest temperature investigated.

In the main figure we see that at intermediate times this
function is peaked at around 0.4\AA~and hardly depends on time. This
observation can be understood by recalling that in this time regime,
the $\beta-$regime, the particles are trapped in the above discussed
cages formed by the surrounding particles.  
Only for times corresponding to the $\alpha-$relaxation this
peak starts to decrease significantly and the function begins to show a
secondary peak, or rather a shoulder, at a distance around 2.6\AA, 
the location of the first
peak in the radial distribution function between two oxygen
atoms~\cite{vollmayr96,horbach_diss}. This secondary peak can be better
seen in a semi-logarithmic plot of this function, which is shown as an
inset of the figure.  Hence we conclude that at low temperatures the
cage for the oxygen atoms breaks up because an atom jumps to the
location of one of its six nearest oxygen neighbors, assuming that no
other oxygen atom is at this position (see also
Ref.~\cite{roux89}).  Note that this type of motion is very different
from the one found in the relaxation of a fragile glass-former
(Lennard-Jones system~\cite{kob_lj}) since in that case the breaking up
of the cage occurs in that the main peak observed in the
$\beta-$relaxation becomes broader and moves to larger distances and no
secondary peak is observed at any time, thus showing that in these
systems no jump processes are present.

The relaxation dynamics of the system can also be studied well by means
of the (incoherent) intermediate scattering function
$F_s^{\alpha}(q,t)$, which is given by a space Fourier-transform of
$G_s^{\alpha}(r,t)$, i.e.

\begin{equation}
F_s^{\alpha}(q,t)=N_{\alpha}^{-1} \sum_{j=1}^{N_{\alpha}} \langle
\exp\left(i\vec{q}\cdot(\vec{r}_j(t)-\vec{r}_j(0))\right)\rangle \quad.
\label{eq5}
\end{equation}

\begin{figure}[t!]
\centerline{\epsfig{file=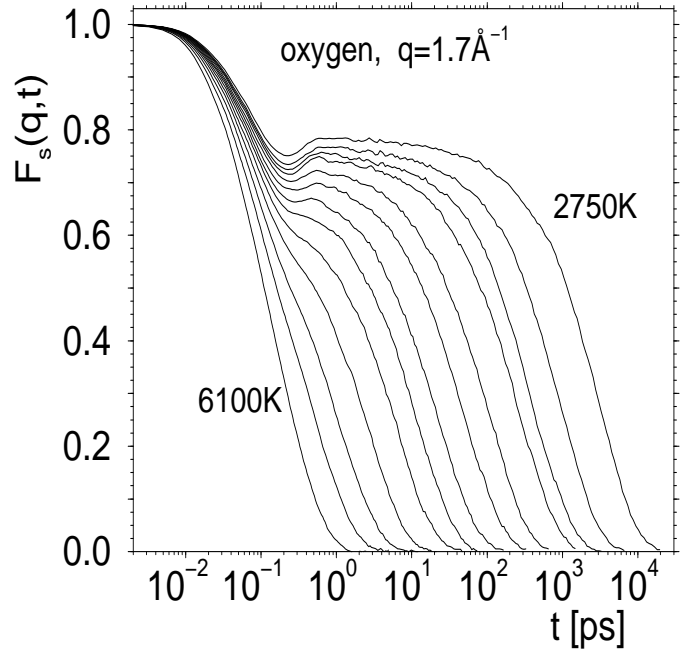,height=3.8in,width=3.8in}}
\vspace{5pt}
\caption{Time dependence of the incoherent intermediate scattering
function for the oxygen atoms for different temperatures.}
\label{fig4}
\end{figure}

The time dependence of $F_s^{\rm O}(q,t)$ is shown in Fig.~\ref{fig4} for all
temperatures investigated. The value of $q$ is 1.7\AA$^{-1}$, 
the location of
the first peak in the static structure factor. We have found that the
correlation functions for different wave-vectors as well as those for
the silicon atoms are qualitatively similar to the one presented
here. The same is true for the coherent intermediate scattering
functions. From the figure we see that with decreasing temperature the
$\alpha-$relaxation time increases quickly and we have found that its
temperature dependence is very similar to the one of the diffusion
constants (see Fig.~\ref{fig2}) in that it shows an Arrhenius
dependence at low temperatures and at high and intermediate
temperatures a dependence which can be fitted well with the power-law
given in Eq.~(\ref{eq3})~\cite{horbach98d}. (It was from this
temperature dependence that the critical temperature $T_c=3330$~K was
determined.) In particular we find that for temperatures above $T_c$,
the product between $\tau$ and $D$ was essentially constant, as it is
predicted by MCT, whereas below $T_c$ the product increases (which is
not in contradiction to the extended version of MCT).

We see that also in $F_s(q,t)$ the cage effect is seen very well in
that at low temperatures the relaxation function shows a plateau.
Before the correlators reach this plateau, they show a pronounced dip, a
feature not present in the dynamics of fragile glass-formers. This dip
is the signature of the so-called boson-peak, a dynamical feature whose
origin is still a matter of debate~\cite{horbach98a,bp_papers}. 

For the time-range of the $\alpha-$relaxation, MCT predicts that the
so-called time-temperature superposition principle (TTSP) 
holds, i.e.~that a time correlation function $\phi(t)$ can be written
as
\begin{equation}
\phi(t)=\Phi(t/\tau(T)) \qquad,
\label{eq6}
\end{equation}
where $\Phi$ is a master function which will depend on $\phi$. This
prediction of the theory can easily be tested by plotting the time
correlation function versus the rescaled time $t/\tau$, which is done
\begin{figure}[t!]
\centerline{\epsfig{file=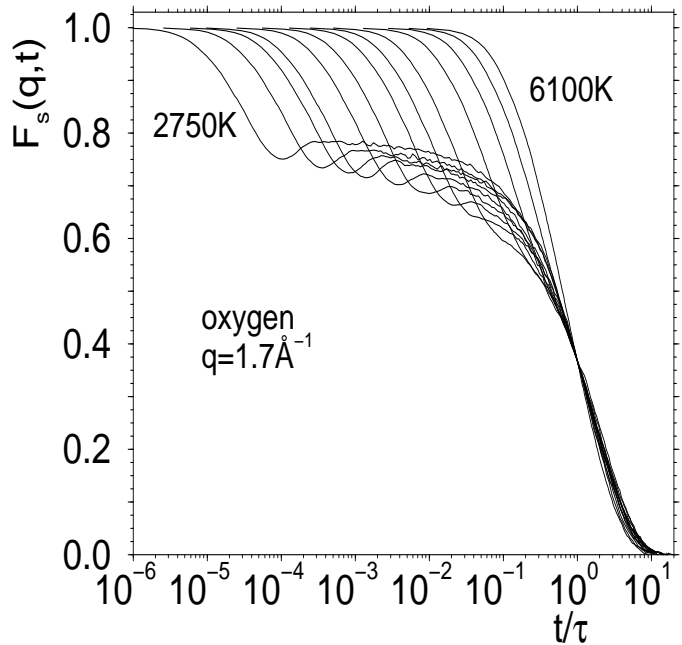,height=3.8in,width=3.8in}}
\vspace{5pt}
\caption{Incoherent intermediate scattering function for the oxygen
atoms as a function of
rescaled time $t/\tau$.}
\label{fig5}
\end{figure}
in Fig.~\ref{fig5} for the intermediate scattering function shown in
Fig.~\ref{fig4}. From the figure we see that in the $\alpha-$relaxation
regime this type of scaling leads indeed to a nice collapse of the
curves for the different temperatures, as predicted by the theory. In
the time regime of the late $\beta-$relaxation, however, strong
deviations from the TTSP are observed, in contrast to what has been
found in many fragile glass formers. A detailed analysis of the curves
has shown~\cite{horbach_diss} that the violation of the TTSP in this
time regime is due to the presence of the boson-peak, since the
intensity of this feature depends significantly on temperature and thus
leads to a strong temperature dependence of the height of the plateau
and hence to a violation of the TTSP in this time window.  However, the
{\it shape} of the curves was found to be essentially independent of
temperature, in agreement with the prediction of
MCT~\cite{horbach_diss}. In particular we have found that in the late
$\beta-$relaxation regime this shape, i.e. the time dependence, is
described very well by the so-called von Schweidler law, i.e.
$F_s(q,t)= f_s^c(q)-h(q) t^b$, where according to MCT the exponent $b$ is
related to the critical exponent $\gamma$ of
Eq.~(\ref{eq3})~\cite{mct}, a relationship which we have found to hold
to within 10\% for the present system.

\begin{figure}[t!]
\centerline{\epsfig{file=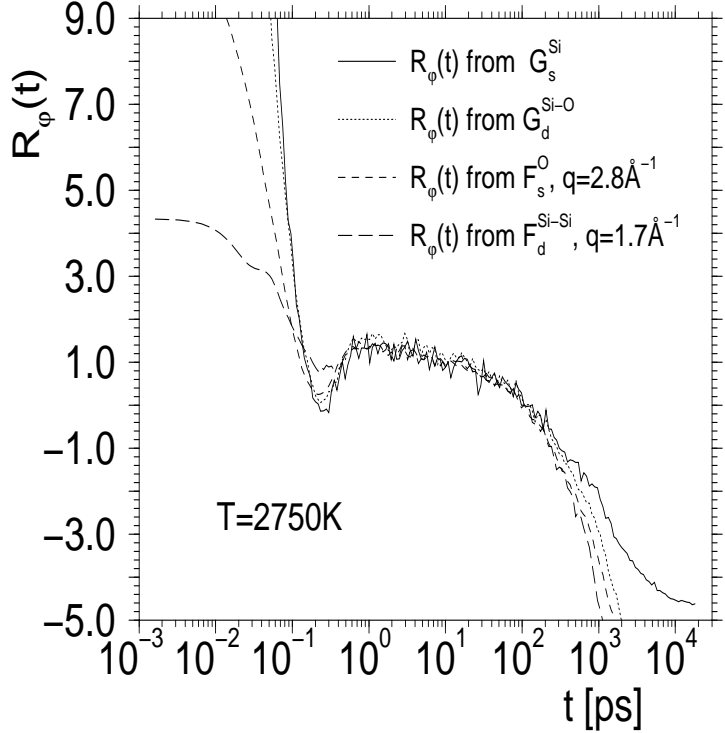,height=3.8in,width=3.8in}}
\vspace{5pt}
\caption{The ratio $R_{\phi}(t)$, see Eq.~(\protect\ref{eq8}), for different
correlation functions $\phi$ as a function of time. $T=2750$~K. See text for
details.}
\label{fig6}
\end{figure}

It is sometimes argued that the ideal version of MCT does not
hold below $T_c$, since real systems will always show ergodicity
restoring hopping processes, and since the details of the extended
theory have not been worked out, the whole theory becomes useless at
low temperatures. In the following we will demonstrate that such an
attitude is overly pessimistic and does not correspond to reality.
Although the extended version of MCT does not make predictions on the
$\alpha-$relaxation regime, it makes useful and testable prediction in
the $\beta-$relaxation regime. In particular the theory predicts that
in this regime the so-called factorization property should hold. This
means that any time correlation function $\phi(t)$ can be written as
follows:
\begin{equation}
\phi(t)=\phi_c+hg(t) \quad.
\label{eq7}
\end{equation}
Here the constant $\phi_c$ is the height of the plateau (see
Fig.~\ref{fig4}), also sometimes called nonergodicity parameter, and
$h$ is a constant which is called critical amplitude. Both quantities
depend on $\phi$. The important point is that the time dependence of
$\phi$ is given by the {\it system universal} function $g(t)$, i.e.~a
function that does not depend on $\phi$. One possibility to test
whether the factorization property holds is to calculate the ratio
\begin{equation}
R_{\phi}(t)=\frac{\phi(t)-\phi(t')}{\phi(t'')-\phi(t')} \quad.
\label{eq8}
\end{equation}
Here $t'$ and $t''\neq t'$ are two arbitrary times {\it in the
$\beta-$relaxation regime}. A trivial calculation shows that the
quantity $R_{\phi}(t)$ will be independent of $\phi$ if $\phi(t)$ can
be written in the form given by Eq.~(\ref{eq7}). In order to test this
prediction of the theory we therefore plot in Fig.~\ref{fig6} the time
dependence of $R_{\phi}$ for different correlators $\phi(t)$ at
$T=2750$~K. In particular we use for $\phi(t)$ the incoherent
intermediate scattering function for
oxygen at $q=2.8$\AA$^{-1}$, the coherent intermediate scattering
function for the Si--Si correlation at $q=1.7$\AA$^{-1}$, the function
$b_{s}(t)=\int_{0}^{0.4 \AA} 4\pi r^2G_s^{\rm Si}(r,t)dr$, and $a_{\rm
Si-O}(t)=\int_{1.32\AA}^{2.35\AA}4\pi r^2\left[G_d^{\rm
Si-O}(r,t)-1\right]\left[g_{\rm Si-O}(r)-1\right]dr$. Here $G_d^{\rm
Si-O}(r,t)$ and $g_{\rm Si-O}(r)$ are the distinct part of the van
Hove correlation function and the radial distribution function between
Si--O atoms, respectively~\cite{hansen86}. (These two last functions
have been proposed by Roux {\it et al.} and have no deep physical
meaning, but are just functions which can be calculated with relatively
high accuracy~\cite{roux89}. The occurring integration boundaries are
related to the location of the minima in the radial distribution
function~\cite{horbach_diss}.)

From the figure we see that in the time scale of the
$\beta$-relaxation regime the different curves collapse nicely onto a
master curve, as it is predicted by MCT. For times outside this
regime no collapse is observed, demonstrating that the existence of a
master curve is a nontrivial feature. We emphasize that the curves
are for a temperature which is considerably below $T_c$ and that, as
demonstrated above by means of the diffusion constants and the van
Hove correlation functions, at this temperature the hopping processes
are very prominent. Thus we conclude that, although not all details
of the extended version of MCT have been worked out, this theory is
indeed able to make interesting and testable predictions also below
$T_c$. Finally we mention that we have found that the factorization
property also holds well at temperatures as high as $T=4000$~K, i.e.
above the critical temperature $T_c=3330$~K, in agreement with the
theory.

\section*{Summary and Discussion}
We have presented the results of a large scale molecular dynamics
computer simulation of a realistic model, the BKS model~\cite{beest90},
of supercooled silica, the prototype of a strong glass-former. Although
we find that the time dependence of the mean squared displacement of a
tagged particle is roughly similar to the one in a fragile
glass-former~\cite{kob_lj}, important differences can be noted, such
as, e.g., an overshoot which is related to the boson-peak as well as
the temperature dependence of the long time behavior. The latter is
related to the temperature dependence of the diffusion constant $D$ and
we find that $D(T)$ shows at low temperatures an Arrhenius behavior
with activation energies that agree very well with experimental values,
thus giving evidence that this silica model is indeed quite realistic.
With increasing temperature significant deviations from the
Arrhenius-law are observed and we find that the temperature dependence
of $D$ can be fitted well with the power-law proposed by MCT with a
critical temperature $T_c$ around 3330~K.  Thus we find that for this
strong glass former $T_c$ is more than twice the experimental glass
transition temperature $T_g$ which is at 1450~K. (We note that in
Ref.~\cite{horbach98d} we have given evidence that the BKS model has an
{\it experimental} glass transition temperature which is close to the
real one.) Thus these results support the idea~\cite{rossler96} that
also for strong glass-formers there is a temperature range in which the
predictions of the ideal version of MCT hold, that, however, for the
case of silica this range is at temperatures which are currently
difficult to access by experiments.

For temperatures below $T_c$ we find, by investigating the self part of
the van Hove correlation function, that the motion of the particles is
strongly influenced by jump like motions of the oxygen atoms, i.e. that
instead of the flow-like motion of the particles proposed by the ideal
version of MCT above $T_c$, it might be more appropriate to think of a
quasi-frozen potential energy landscape in which the dynamics is mainly
given by the hopping of the particles~\cite{goldstein69}. That the
presence of these strong hopping processes does not exclude the
possibility that MCT makes useful predictions also at low temperatures,
is demonstrated by our investigation of the factorization property~[see
Eq.~(\ref{eq7})]. We show that this property holds very well even at
temperatures at which the dynamics of the system is strongly influenced
by hopping processes. Furthermore we have found~\cite{horbach98d} that
the master curve obtained, see Fig.~\ref{fig6}, is in the late part
given by the von Schweidler law with an exponent $b$, which fulfills the
relation proposed by MCT between $b$ and the critical exponent $\gamma$
of the relaxation times~\cite{mct}.

Finally we have also investigated the time and temperature dependence
of the incoherent intermediate scattering function. We find that the
time-temperature superposition principle holds well in the late
$\alpha-$relaxation regime that, however, quite large deviations are
observed in the $\beta-$relaxation regime. The reason for these
differences is the boson-peak, which leads to a temperature dependence
of the height of the plateau. The form of the curves, however, is
indeed independent of temperature, as can be seen from the
factorization property.

Thus we can summarize by saying that many aspects of the dynamics of
this strong glass-former can be understood by means of MCT, although
the agreement between the {\it ideal} version of the theory and the
results of the simulation is less satisfying as it has been found to be
for fragile glass-formers.  In particular the range over which the
power-law for the diffusion constants can be observed is relatively
small and the strong influence of the boson-peak onto the
$\beta-$relaxation leads to a breakdown of some, but not all, of the
predictions of the theory in this time window.  Nevertheless, the {\it
extended} version of the theory is able to describe correctly some
aspects of the dynamics also at low temperature, a result which should
encourage to work out this version of the theory in more detail.

\section*{ACKNOWLEDGMENTS}
This work was supported by BMBF Project 03~N~8008~C and by SFB 262/D1
of the Deutsche Forschungsgemeinschaft.  We also thank the HLRZ
J\"ulich and the RUS in Stuttgart for a generous grant of computer time
on the T3E.

\end{document}